\documentclass[journal=jacsat,manuscript=article,layout=twocolumn]{achemso}
\usepackage{chemformula} 
\usepackage{color,soul}
\usepackage[T1]{fontenc} 
\usepackage{amsmath,amssymb}

\usepackage{tabularx}
\usepackage{array}
\usepackage{dcolumn}
\usepackage{setspace}

\author{Andrei Kudriashov}
\email{andrei.kudriashov.97@gmail.com}
\affiliation{Advanced mesoscience and nanotechnology centre, Moscow Institute of Physics and Technology, 141700 Dolgoprudny, Russia}

\author{Ian Babich}
\affiliation{Advanced mesoscience and nanotechnology centre, Moscow Institute of Physics and Technology, 141700 Dolgoprudny, Russia}

\author{Razmik A. Hovhannisyan}
\affiliation{Advanced mesoscience and nanotechnology centre, Moscow Institute of Physics and Technology, 141700 Dolgoprudny, Russia}

\author{Andrey G. Shishkin} 
\author{Sergei N. Kozlov}
\affiliation{Advanced mesoscience and nanotechnology centre, Moscow Institute of Physics and Technology, 141700 Dolgoprudny, Russia}
\affiliation{Fundamental Physical and Chemical Engineering Department, MSU, 119991 Moscow, Russia}
\affiliation{Dukhov Research Institute of Automatics (VNIIA), Moscow 127055, Russia}

\author{Alexander Fedorov} 
\affiliation{Leibniz Institute for Solid State and Materials Research, P.O. Box 270116, D-01171 Dresden, Germany}
\author{Denis V. Vyalikh}
\affiliation{Donostia International Physics Center (DIPC), 20018 Donostia-San Sebastian, Basque Country, Spain}
\alsoaffiliation{IKERBASQUE, Basque Foundation for Science, Bilbao, 48011 Spain}

\author{Ekaterina Khestanova}
\affiliation{Department of Physics and Engineering, ITMO University, Saint Petersburg, Russia}

\author{Mikhail Yu. Kupriyanov}
\affiliation{Skobeltsyn Institute of Nuclear Physics, MSU, Moscow 119991, Russia}

\author{Vasily S. Stolyarov}
\affiliation{Advanced mesoscience and nanotechnology centre, Moscow Institute of Physics and Technology, 141700 Dolgoprudny, Russia}
\alsoaffiliation{Dukhov Research Institute of Automatics (VNIIA), Moscow 127055, Russia}
\alsoaffiliation{National University of Science and Technology MISIS, 119049 Moscow, Russia}

\title[An \textsf{achemso} demo]{Revealing intrinsic superconductivity of the Nb/BiSbTe$_2$Se interface}

\begin{document}
\begin{figure}[ht!]
\begin{center}
\includegraphics[width=8.6cm]{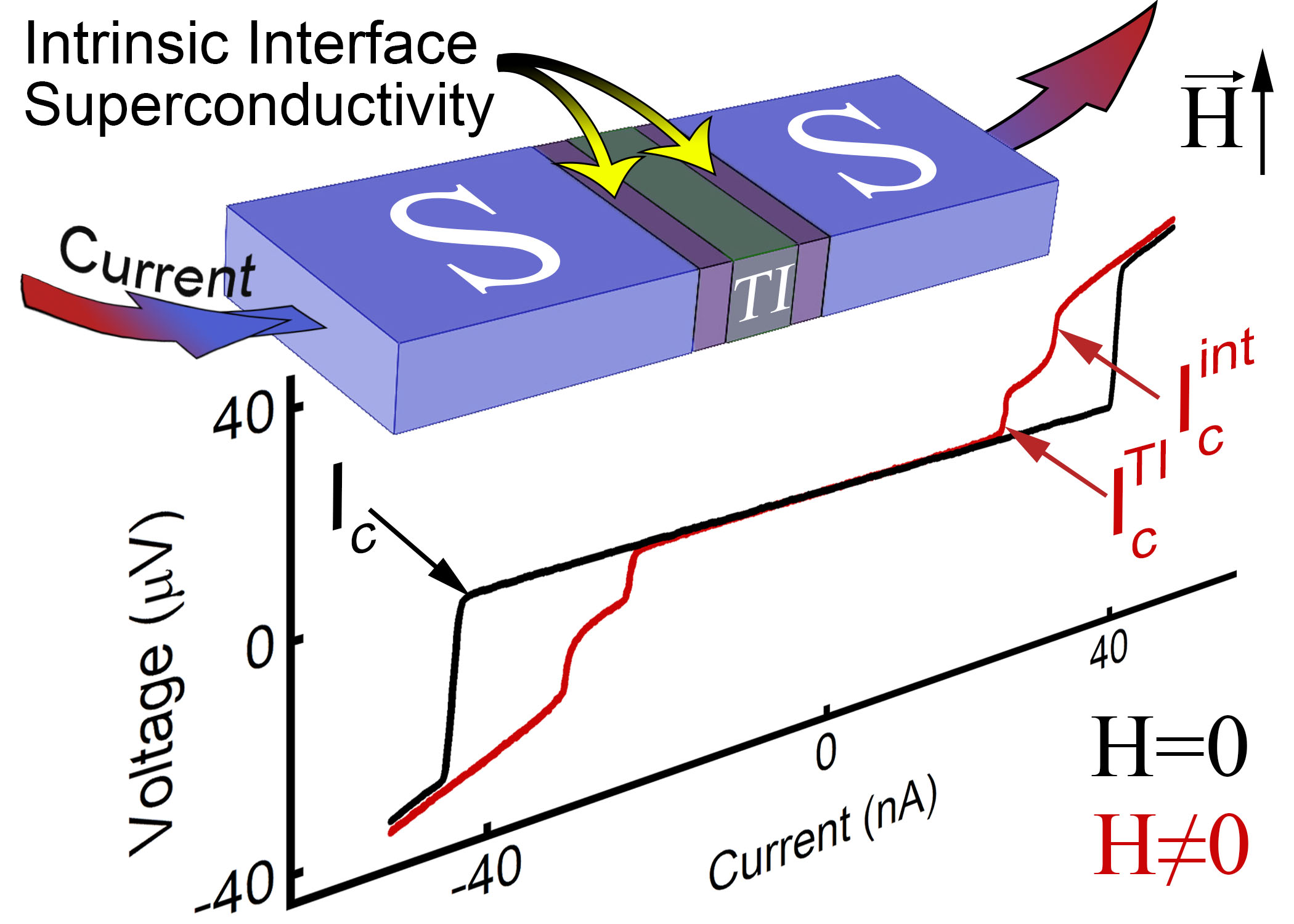} 
\end{center}
\end{figure}

\begin{abstract}
Typically, topological superconductivity is reachable via proximity effect by a direct deposition of superconductor (S) on top of a topological insulator (TI) surface. 
Here we observed and analysed the double critical current in the Josephson junctions based on the topological insulator in the fabricated planar Superconducting Quantum Interference Device.
By measuring critical currents as a function of temperature and magnetic field, we show that the second critical current stems from the intrinsic superconductivity of the S/TI interface, which is supported by the modified Resistively  Shunted Junction model and Transmission Electron Microscopy studies. 
This complex structure of the interface should be taken into account when technological process involves Ar-plasma cleaning. 
\end{abstract}

\section{Introduction}
The topology of electronic band structure plays a central role in explaining different physical phenomena in condensed matter systems\cite{narang2021,bradlyn2017,vergniory2019}. 
The non-triviality of the band structure of some materials leads to the formation of new states of matter, such as topological insulators (TIs) \cite{hasan2010}, Dirac or Weyl semimetals \cite{yan2017,wang2017}.
Furthermore, such non-trivial bands have the potential to induce topological superconductivity via the proximity effect \cite{fu2008, sato2009, tanaka2009, qi2011}.
Therefore, the interface between topological insulator and superconductor plays an extremely important role in studying topological superconductivity\cite{stolyarov2022,stolyarov2020,ji20181}.
Since the superconducting order parameter is sensitive to microscale structures around electrodes and their fabrication process, it is crucial to evaluate both the transport properties of superconducting junctions and their local structure to establish topological superconductivity.

In this work, we demonstrate the double critical current in the Superconducting Quantum Interference Device (SQUID), fabricated by sputtering of Nb on top of the Fe-doped $\text{Bi}\text{Sb}\text{Te}_{2}\text{Se}$ flake. 
By analyzing those critical currents as a function of temperature and magnetic field, we show that the second critical current stems from the intrinsic superconductivity of the S/TI interface, which is supported by the Restively Shunted Junction (RSJ) model and Transmission Electron Microscopy (TEM) imaging.
Our deep insight into the interface of the Bi-based TI junction provides a future qualitative analysis of emergent unconventional SC states and an avoidance strategy for achieving a unitary topological superconducting state for further TI josepshonics. 

\begin{figure}[ht!]
\begin{center}
\includegraphics[width=8.6cm]{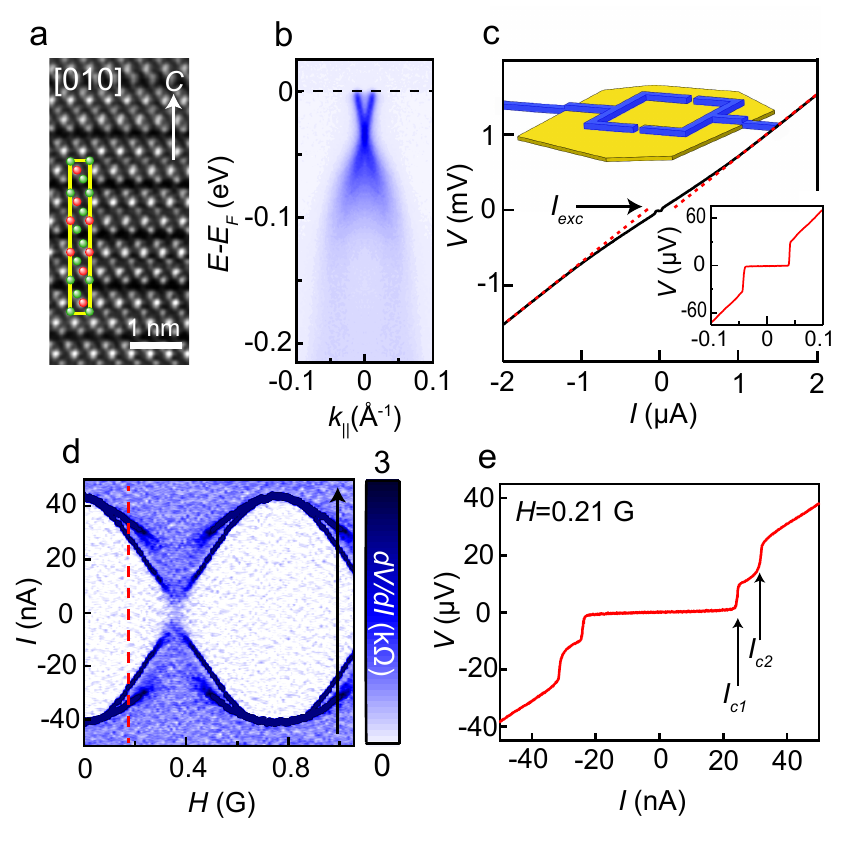} 
\caption{
(a) HAADF-STEM image of atomic planes perpendicular to [010] direction of Fe-BSTS crystal.
Inset shows the schematics of one unit cell.
Green circles are Se/Te atoms, red ones - Bi/Sb. 
(b) ARPES data of Fe-BSTS taken at $T=1.4$ K. 
(c) $I(V)$ characteristic of the SQUID taken at $H=0$ G and $T=26$ mK. 
Red dashed lines are linear fits of the high-voltage regions. 
Top inset shows the schematics of the
device. 
Bottom right inset shows the $I(V)$ characteristic of the device in low current region. 
(d) Color map of differential resistance $dV/dI$ as a function of current and external magnetic field at $T=13$ mK. 
Arrow represents the direction of current sweep. 
(e) $I(V)$ characteristic taken at $H=0.21$ G (red dashed line in the color map) and $T=26$ mK.}

\label{Fig.1}
\end{center}
\end{figure}

\begin{figure*}
\begin{center}
\includegraphics[width=17cm]{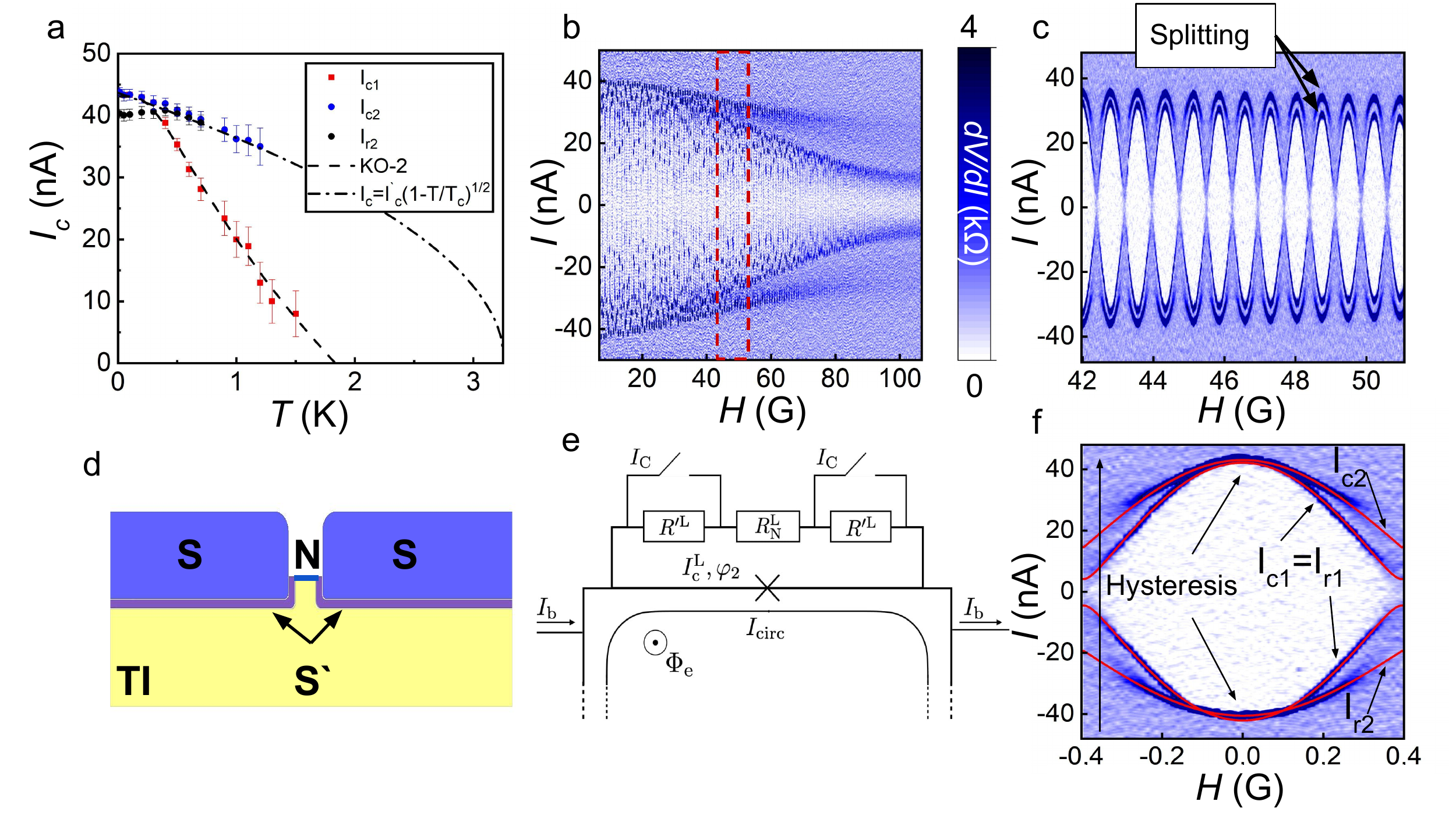} 
\caption{(a) Temperature dependence of critical and retrapping currents. The dashed lines are the fitting lines. See the main text for details.
(b) Color map of differential resistance $dV/dI$ as a function of current and external magnetic field at $T=20$ mK. 
(c) Color map of differential resistance $dV/dI$ as a function of current and external magnetic field at $T=20$ mK of the region, denoted in (b) by red rectangle. 
(d) Schematics of the single Josephson junction based on TI. Blue regions correspond to Nb (S), purple - extra superconducting layer ($S^{\prime}$), yellow - topological insulator flake (TI) with surface state (N). 
(e) RSJ model of the single Josephson junction, taking into account extra superconducting region $S^{\prime}$. 
(f) Color map of differential resistance $dV/dI$ as a function of current
and external magnetic field at $T=13$ mK. The red lines are the fits of $I_{c1}$ and $I_{c2}$ by the modified RSJ model. 
Arrow represents the direction of the current sweep.}
\label{Fig.2}
\end{center}
\end{figure*}

\section{Results}
We studied Fe-doped 3D topological insulator $\text{Bi}\text{Sb}\text{Te}_{2}\text{Se}$ (Fe-BSTS), synthesized by the modified Bridgman method \cite{Rikizo2019}. 
High-resolution high angle annular dark field scanning transmission electron microscopy (HAADF-STEM) image of the Fe-BSTS is shown in Figure \ref{Fig.1}a. 
The electron diffraction image corresponds to the rhombohedral crystal structure with lattice constants $a=4.2$ \AA $ $  and $c=29.7$ \AA, which is consistent with previous measurements of similar TIs \cite{ko2013, abou2008, Rikizo2019}.
Due to the optimized composition, the Fermi level in this material is located within the bulk band gap and 30 meV higher than the Dirac point, according to the angle-resolved photoemission spectroscopy (ARPES), which is shown in Figure \ref{Fig.1}b.
As a result, the low concentration of the bulk carriers leads to the surface dominated transport at low temperature \cite{yano2021}. 

Thin flakes of Fe-BSTS were mechanically exfoliated from the bulk material, transferred to the $\text{Si}/\text{SiO}_2$ substrate, and then SQUID on top of the flake was fabricated using a standard technological process (See Section 1 of the Supplemental Materials for details).
Schematic drawing of the final device is shown in the inset of the Figure \ref{Fig.1}c.

$I(V)$ characteristic at zero magnetic field and base temperature of our dilution refrigerator is shown in Figure \ref{Fig.1}c by the black line. 
The bottom right inset of Figure \ref{Fig.1}c is zoom of $I(V)$ curve in the range from -0.1 $\mu A$ to 0.1 $\mu A$. 
It clearly demonstrates the single critical current, which is typical for Josephson junctions. 
$I(V)$ curve shows nonlinear behaviour when current is higher than the critical current, indicating the presence of excess current.

The device is showing very pronounced oscillations of the critical current in the external magnetic field perpendicular to the surface of the loop (Figure \ref{Fig.1}d).
The period of these oscillations is consistent with the flux quantum $\Phi_0=hc/2e$ per loop area.

Distinctive feature of this device, which can be seen in Figure \ref{Fig.1}d is what we call "second critical current". 
When the external magnetic field corresponds to an integer number of flux quantum, there is only one step in the $I(V)$ characteristics, which is identified as critical current.
Nevertheless, when the external magnetic field is around $\Phi_0/4$, two clear steps in the $I(V)$ characteristics can be seen, which is presented in the Figure \ref{Fig.1}e. 
We determinate the first step as $I_{c1}$ and the second step as $I_{c2}$.

\begin{figure*}
\begin{center}
\includegraphics[width=16cm]{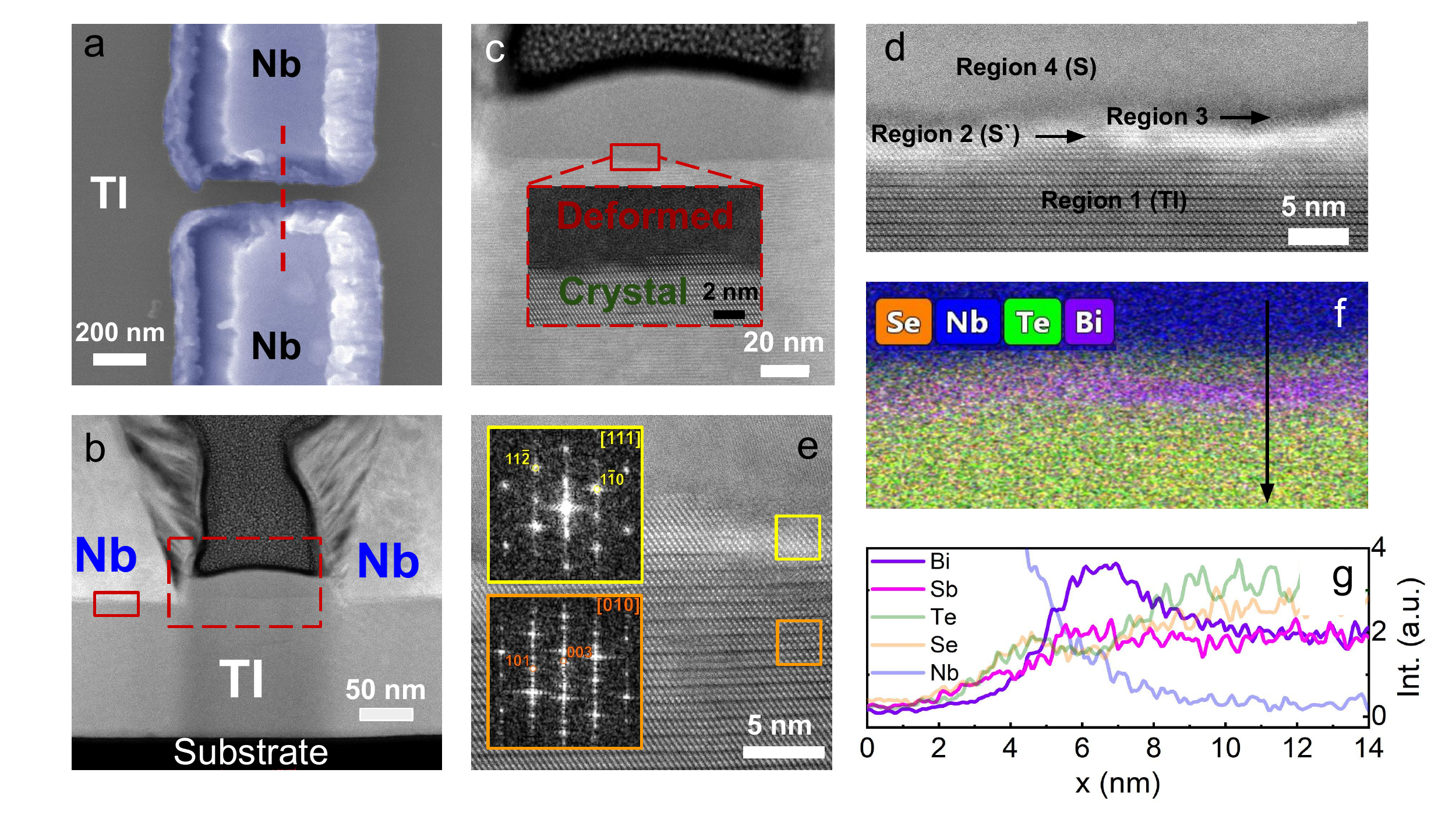} 
\caption{(a) SEM image of the Josephson junction, placed on top of the Fe-BSTS flake. 
(b) TEM image of the cross section, pictured in the (a) by red dashed line. 
(c) and (d) show the high resolution HAADF-STEM image of the regions, pictured in (b) by red dashed rectangle and small red rectangle, respectively. 
(e) FFT of two regions, corresponding to TI (orange) and $S^{\prime}$ (yellow). 
(f) EDX map of the area shown in (d). 
(g) Profile of the elements distribution, subtracted from line, shown in the (f) by the arrow.
}
\label{Fig.3}
\end{center}
\end{figure*}

The temperature dependence of critical current $I_{c1}$, critical current $I_{c2}$ and retrapping current $I_{r2}$ at zero magnetic field is shown in Figure \ref{Fig.2}a by red rectangles, blue circles and black circles, respectively. The hysteresis of the I(V) characteristic is clearly seen in the Figure \ref{Fig.1}d and Figure S3 of the Supplemental Materials. 
The black dashed line is the fit of $I_{c1}(T)$ by the Kulik-Omelyanchuk model in the clean limit (KO-2) \cite{golubov2004}. 
For details of this model and fitting procedure, see Section 4 of Supplemental Materials and Ref. \cite{stolyarov2020}.

At large magnetic fields, instead of the standard modulation of the critical current by the Fraunhofer dependence, a monotonic decrease of envelope curve of the critical current is observed (see Figure \ref{Fig.2}b and Figure S4). 
It suggests that the magnetic field acts as a pair breaking mechanism that suppresses the superconductivity in the surface states \cite{PairBreaking}, making two critical currents split completely, as shown in Figure \ref{Fig.2}c. 
This indicates that the second critical current $I_{c2}$ comes from the properties of the single Josephson junction. Moreover, we have fabricated single Josephson junctions, which also demonstrate this feature (see Section 3 of Supplemental Material for details).

\section{Discussion}
Additional features on the I(V) characteristic are frequently observed in Josephson junctions based on topological insulators \cite{williams2012,UnderNb}.
There could be different explanations, such as Fiske steps~\cite{barone1982physics}, analogues of Shapiro steps\cite{barone1982physics,pribiag2015edge}, McMillan-Rowell\cite{shi2015mcmillan} or Tomasch~\cite{tomasch1965geometrical} resonances. 
We strongly believe that neither of these explanations fits our experimental results. 
First of all, our Josephson junctions are fairly overdamped and resonances such as Fiske can’t be observed in our system. 
Moreover, Shapiro like resonances with outer resonant system (sample holder, for example) can not explain the hysteresis of the $I_{c2}$ we observed, as discussed bellow. 
Additionally, single Josephson junctions were measured in the same sample holder and they show different characteristic voltage, excluding the possibility of unwanted resonances in the sample holder.
McMillan-Rowell or Tomasch resonances are also not the case of our setup as they appear in underdamped JJs with geometrical resonances of quasiparticle or superconducting Andreev states.

Generally, a planar Josephson junction can be described as an $SS^{\prime}NS^{\prime}S$ structure~\cite{kunakova2019}.
Here, $S$ corresponds to Nb, $S^{\prime}$ is the region of TI under the superconducting electrode, and N is the surface state of the TI, as shown in the Figure \ref{Fig.2}d).
Another explanation for the second feature on the I(V) characteristic is that  $I_{c2}$ is related to the region $S^{\prime}$ located between Nb and TI.

We model our SQUID with an extra superconducting layer $S^{\prime}$ underneath the Nb contacts (Figure \ref{Fig.2}d) with a modified RSJ model. 
Its equivalent circuit is shown in Figure \ref{Fig.2}e. 
For bias currents below $I_{c2}$, a single JJ in the SQUID acts as an $SS^{\prime}NS^{\prime}S$. 
When the bias current satisfies the condition $|I_b|/2+|I_{circ}|>I_c'$ (see Section 5 of the Supplemental Materials for more details), the JJ acts as an $SN^{\prime}NN^{\prime}S$. 
Here $I_c', I_b, I_{circ}$ are the $S^{\prime}$-layer critical current, bias current, and the current circulating in the loop, respectively. 
The switching from $SS^{\prime}NS^{\prime}S$ to $SN^{\prime}NN^{\prime}S$ leads to the second voltage step, which can be described as an appearance of $N^{\prime}$-layer resistance $R'$ in the shunt.

Numerical simulations predict that $I_{c2}$ depends on magnetic field as $A+B|\cos(\pi\Phi/\Phi_0)|$, which nicely fits the experimental data (Figure \ref{Fig.2}f). 
Here, A and B are the fit parameters. The $I_{c1}$ behaviour is fitted as an asymmetric SQUID with asymmetry coefficient $\alpha=\frac{I_{c}^{L}-I_{c}^{R}}{I_{c}^{L}+I_{c}^{R}}=0.1$, where $I_{c}^{L}$ and $I_{c}^{R}$ are the critical currents of the left and the right Josephson contacts in the SQUID.

Typically, the retrapping current is different from the critical current in SNS junctions because of the overheating of the N region \cite{courtois2008}.
To describe the fact that $I_{c1}$ is non-hysteretic but $I_{c2}$ is (see Figures \ref{Fig.2}a,f and Section 2 of the Supplemental Materials), we added Joule heating in the model by thermally coupling $S^{\prime}$ interfaces, N region, S electrode, and the TI flake \cite{biswas2018}. 
Details of this model could be found in Section 5 of the Supplemental Materials.
When the bias current reaches the retrapping current, the $N^{\prime}$-layer switches to an $S^{\prime}$ and $R'$ drops to 0. 
The model suggests that $R'\sim R_N$, therefore, the produced heat significantly decreases. 
The excess heat is being transferred to the flake through metallic surface states, thus the Josephson junction region quickly cools down to the bath temperature. 
As a result, the SQUID does not exhibit the switching, corresponding to $I_{c1}$, because $I_{r2}<I_{c1}$. 
That is why we observe that $I_{r2}$ limits $I_{c1}$ near the integer flux quantum values (bottom curves in the Figure \ref{Fig.2}f). 

In order to describe the blurriness of the second critical current around $\Phi=\Phi_0/2$ (see Figure \ref{Fig.1}d), we take into account the inhomogeneity of the $S^{\prime}$ layer and treat the $S^{\prime}$ as a combination of layers with slightly different $I_c'$. 
Indeed, this approach provides a nonzero width of $I_{c2}$, if $R_{N}^{L\prime}> R_{N} > R_{N}^{R\prime}$ inequality holds (see Section 5 of the Supplemental Materials). 
However, it does not lead to the full evanescence of $I_{c2}$ at half flux quantum. Another possible explanation of the blurriness effect is the interplay between the SQUID loop and a superconducting loop formed by $S^{\prime}$ region. The difference in London penetration depths and geometric inductance of the loops may result in a peculiar behaviour of screening currents that pass through the weak link region and are not considered by the RSJ model.  

To validate the presence of the $S^{\prime}$ region and to study the $SS^{\prime}$ and $S^{\prime}N$ interfaces, we performed TEM study of the interface between Nb and Fe-BSTS. 
We cut one of the Josephson junctions, as shown in Figure \ref{Fig.3}a along the red line. 
The corresponding cross section is shown in Figure \ref{Fig.3}b. 
Since we have conducted the TEM study after the transport measurements (and, therefore, thermal cycling) we have observed the same strain-induced buckling feature, as in Ref.\cite{charpentier2017}, which is indicated in Figure \ref{Fig.3}b by a large rectangle. 
Interestingly, the interface between strained and unstrained regions is very sharp, as shown in Figure \ref{Fig.3}c. 
Unstrained region demonstrates well-ordered structure, while the strained region is so deformed that there is no order seen. 
Nevertheless, as in Ref.\cite{charpentier2017}, this feature was formed after the worm-up and therefore did not affect our results.

There are regions of specific contrast between the TI layers (region 1) and Nb contacts (region 4) (see Figure \ref{Fig.3}d). 
One of it looks darker (region 3) and does not have a periodic structure, while the other (region 2), on the contrary, has a brighter contrast and demonstrates local ordering. 
A redistribution of the intensity in the region 2 can be noticed. 
In the ordered structure of $BiSbTe_2Se$, within one quintuple layer, two bright (Bi/Sb) and three less bright (Se/Te) points are observed (see Figure \ref{Fig.3}e and Figure \ref{Fig.1}a), while in region 2 the intensity of all atomic columns is more or less the same, which should indicate the averaging of the atomic mass of all atomic columns, that is, mixing of atoms in all or almost all crystallographic positions and a change in the symmetry of the structure. 
Indeed, the Fourier transform (FFT) taken from the ordered TI layers (see Figure \ref{Fig.3}e, orange inset) and from the region 2 (see Figure \ref{Fig.3}e, yellow inset) are different. 
The FFT of the disordered region lacks reflections corresponding to long distances of ~ 29.7 \AA$ $, and the observed image can be interpreted as the [111] zone of an I-centered tetragonal cell with parameters a=4.9 \AA$ $  and c=6.2 \AA. 

The mixed EDX maps (see Figure \ref{Fig.3}f) show that region 2 is enriched in Bi and Sb, although antimony is not shown on the mixed EDX map for ease of interpretation. 
The same trend is clearly visible on the linear profile of the EDX signal in Figure \ref{Fig.3}g. 
Indeed, the atomic ratio  Bi:Sb:Te:Se in the region 2 corresponds to the composition $\text{Bi}_{2.3}\text{Sb}_{1.2}\text{Te}_{0.9}\text{Se}_{0.6}$, significantly differs from the composition $\text{Bi}_{1.06}\text{Sb}_{0.99}\text{Te}_{1.73}\text{Se}_{1.22}$ obtained for ordered TI layers. 
At the same time, the composition of the amorphous diffuse layer (region 3) is equal to $\text{Bi}_{1.1}\text{Sb}_{1.1}\text{Te}_{1.6}\text{Se}_{1.2}$ and is very close to the nominal composition observed in the TI. Thus, the bright contrast in region 2 is primarily associated with the enrichment of this region with the heaviest bismuth compared to other elements.

We associate region 2 with $S^{\prime}$.
This region can be either proximitized by $S$, or it can has intrinsic superconductivity. 
Usually, superconducting states just below SC electrodes stem from the superconducting proximity effect. However, TEM results suggest other origins, such as intermixing of the elements \cite{schuffelgen2017bad, bai2020}, strain \cite{he2020, he2019}, and other composition issues. 
For example, $\text{Bi}_{1-x}\text{Sb}_x$ alloys are known to be superconducting \cite{Zally1971,Kasumov1996} and the region 2 is enriched in Bi  and  Sb.
In addition, it is known that tetradymite topological insulators becomes superconducting under pressure. 
The superconducting phase transition is accompanied with two structural phase transitions, according to \cite{cai2018}. 
In our case we also observed the tetragonal phase, which is in qualitative agreement with structure, found in Ref.\cite{cai2018}.
Therefore, we suggest that $S^{\prime}$ has intrinsic (not proximitized by Nb) superconductivity. 
Another argument for the intrinsic superconductivity of $S^{\prime}$ comes from the fact that $I_{c2}$ has a linear dependence on external magnetic field (see Figure \ref{Fig.2}b and Section 2 of the Supplemental Materials), which is typical for thin films of superconductors\cite{osti}.
$I_{c2}(T)$ also support this hypothesis. 
It was shown that if $\Delta_{Nb}>>\Delta_{S^{\prime}}$ (i.e. $T_c^{Nb}>>T_c^{S^{\prime}}$), the temperature dependence of the critical current of the $SS^{\prime}$ interface in the KO-1 model is determined by the temperature dependence of $\Delta_{S^{\prime}}$, which can be approximated by the square root~\cite{golubov2004,cherkez2014proximity}. 
In our case, $I_{c2}(T)$ is well fitted by the square root dependence (see Figure \ref{Fig.2}a and the Section 2 of the Supplemental Material), which is another argument for intrinsic superconductivity of $S^{\prime}$.
Therefore, in our case, the second step in the $I(V)$ curve is associated with breaking of the superconductivity in the $S^{\prime}$ region. 

We believe that the $S^{\prime}$ region is formed during the Ar-plasma etching, performed before the Nb deposition. 
It explains why the Region 2 is enriched in Bi and Sb, since it is known that Se and Te are more prone for sublimation, comparing with Bi and Sb. 
In addition, plasma treatment can change the crystal structure of the material, as shown, for example, in Ref.\cite{cai2022}.
In contrast, the direct deposition technique just after cleaving crystals allows to obtain a clean interface without an additional phase, akin to Ref.\cite{schuffelgen2017}. 
Nevertheless, an additional study is required to understand the exact impact of the Ar-plasma to surface of topological insulator.

\section{Conclusion}
Thus, we have demonstrated the existence of two critical currents in SQUID based on Fe-doped TI BiSbTe$_2$Se.
We analyzed the dependencies of these critical currents on temperature and magnetic field and carried out structural and elemental analyses of the Nb interface with Fe-doped TI BiSbTe$_2$Se.
The conducted study shows that the detected effect can be explained by the presence of intrinsic superconductivity in the transition layer between Nb and the Fe-doped TI BiSbTe$_2$Se.
Apparently, this layer is formed during etching of the TI surface and subsequent deposition of Nb.
If so, we are raising a serious concern of this common approach for device fabrication.
We claim, that even in the case of the high S/TI interface transparency, determined from the $I_c(T)$ dependence, the interface may have complex structure, which can adversely affects the physics of artificially designed topological superconductors, including Majorana-related effects. 
This complex structure of the interface should be taken into account when technological process involves Ar-plasma cleaning. 
If atomically flat and clean interface is needed, we suggest to use determenistic transfer inside the glove box \cite{yabuki2016, wang2013, novoselov2016, rooney2017, chong2018}, \textit{in situ} methods\cite{schuffelgen2017,stolyarov2021superconducting}, or other technological processes.

\begin{acknowledgement}
It is a pleasure to thank A. Abramov for the help with the device fabrication; Rikizo Yano, Hishiro T. Hirose, Tsuyoshi Tanda, Takao Sasagawa and Satoshi Kashiwaya for useful discussions.
This work was supported by the RSF 21-72-00140.
This work was performed using e-beam lithography of MIPT Shared Facilities Center.  ARPES studies supported by the Ministry of Science and Higher Education of the Russian Federation (No. FSMG-2021-0005).
TEM study was conducted at Advanced Imaging Core Facility of Skoltech.
Crystal growth was supported by CRP in MSL-Tokyo Tech, Japan.

\end{acknowledgement}

\textbf{Author contribution} 
V.S.S. conceived the project and supervised the experiments, 
A.F. and D.V.V. performed the ARPES measurements,
E.K. did mechanical exfoliation,
A.G.S. and A.K. realized e-beam lithography,
V.S.S performed the deposition of the Nb film,
A.K. conducted the transport measurements,
M.Y.K.,  V.S.S. and R.A.H. provided the explanation of the observed effects,
I.B. and R.A.H. constructed the RSJ-model,
I.B. did numerical modeling,
S.N.K did the fitting by the KO-2 model,
A.K. wrote the manuscript with the contributions from other authors.

\textbf{Competing Interests}
The Authors declare no Competing Financial or Non-Financial Interests.

\textbf{Data Availability}
The data that support the findings of this study are available from the corresponding
authors upon reasonable request. 




\bibliography{bibliography}

\end{document}


\renewcommand{\thetable}{S\arabic{table}}

	\begin{center}
		\Large
		Supplemental material
		
		Revealing intrinsic superconductivity of the Nb/BiSbTe$_2$Se interface
	\end{center}
	
\section{Main device fabrication and Characterization}
The $Si/SiO_2$ substrate was cleaned using ultrasonication in Acetone for 10 min, then in IPA for 5 min and in Deionized Water for 3 min. 
After that, water was blew off the substrate by nitrogen gun.
Finally, the substrate was etched in $O_2$ plasma for 10 min using Diener Atto plasma cleaner.  

The flakes of Fe-BSTS were produced by mechanical exfoliation.
First, bulk single crystal was placed on the blue tape, slightly pressed by tweezers and then removed. 
Parts of the material transferred to the blue tape because of the adhesion to the tape. 
Second, the blue tape with Fe-BSTS crystals on it was pilled-off few times. 
Finally, the blue tape with flakes on it was placed on top of the clean $Si/SiO_2$ substrate, slightly pressed by finger and then relatively quickly removed. 
As a result, part of the nano-crystals was cleaved and remained on the substrate.

The substrate with flakes on it was studied by Optical Microscope, flakes with appropriate size and shape were selected and their coordinate was determined by special coordinate grid, which was fabricated before the substrate cleaning by standard e-beam lithography and e-beam evaporation of 100 nm of $Ti$, followed by lift-off. 

\begin{figure}[ht!]
\begin{center}
\includegraphics[width=
\textwidth]{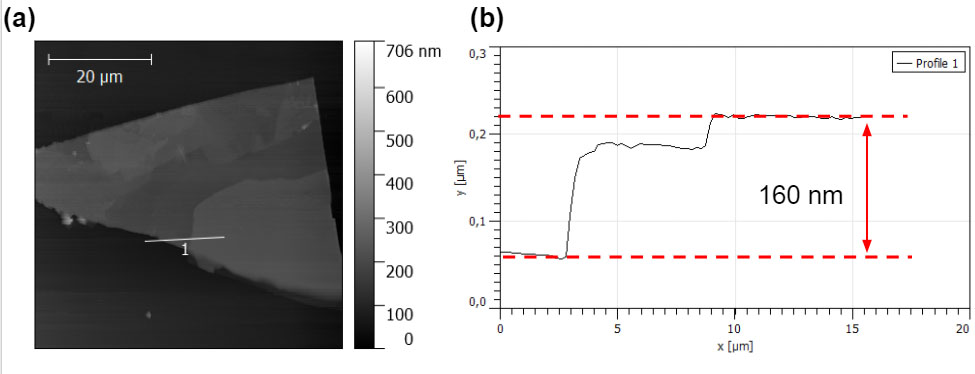} 
\caption{(a) AFM-Image of the Fe-BSTS flake. (b) Height profile along the line 1 at (a)}
\label{AFM}
\end{center}
\end{figure}

Then, selected flakes were studied by Atomic Force Microscopy (AFM) in order to determine the flake thickness and find the flat surface. 
The AFM image and the height profile along the line 1 is shown in Fig.\ref{AFM} (a) and (b), respectively.

\begin{figure}[ht!]
\begin{center}
\includegraphics[width=
\textwidth]{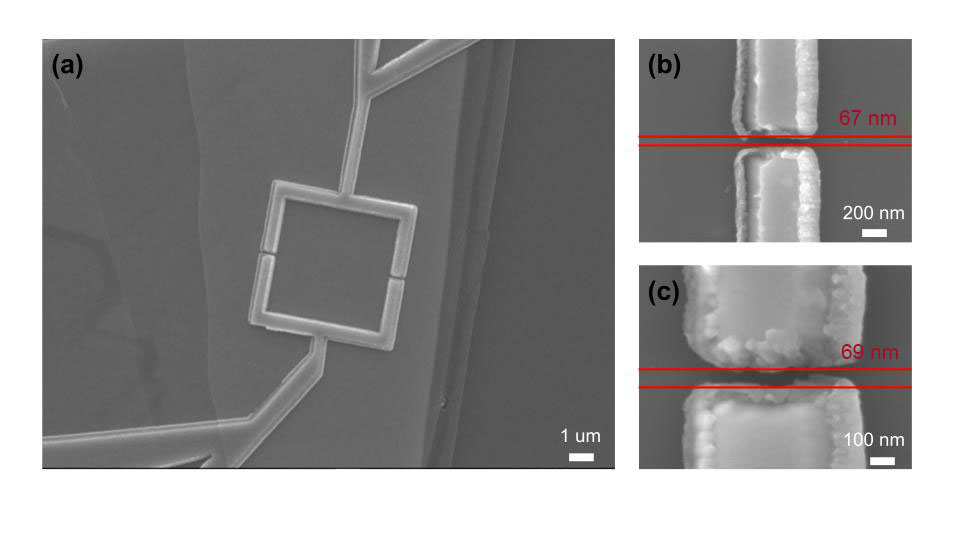} 
\caption{(a) SEM-Image of the final device. (b) and (c) shows the left and the right JJ, respectively.}
\label{SEM}
\end{center}
\end{figure}

To form the SQUID we used e-beam lithography. Double layer PMAA A4 resist was used. 
Each layer was spin coated for 30 sec at 6000 rpm and then annealed at 50 degree Celsius in laboratory oven for 10 hours in order to minimize the heating of the flake. 
After the lithography the resist was developed in MIBK:IPA (1:3) solution for 1.5 min. 
Then, the sample was loaded into the UHV chamber of the metal deposition system, and 250 nm of Nb was deposited by magnetron sputtering. 
Prior to the deposition, 10 sec of Ar plasma etching was performed in order to remove the oxide layer from the flake underneath the Nb contacts. 
Finaly, the sample was left over the night in Acetone in order to lift-off the metal film. 
The SEM image of the final device is shown in Fig.\ref{SEM}.

\section{Details of transport measurements}
We used dilution refrigerator BlueFors LD-250, equipped with 9T superconducting solenoid.
Yokogawa GS-200 was used to charge the solenoid. In order to measure I-V characteristic we used Keithley 6220 current source and Keithley 2182 nanovoltmeter.
Hand-made $2^\text{nd}$ order RC-filter was placed at the cold plate of the cryostat. It should be noted, that the base temperature of our cryostat changed a little over time because of the inappropriate grounding and ground loops issue. That is the reason why the data, presented in the main text, has slightly different temperature (14 mK, 20 mK, 26 mK). 

\begin{figure}[H]
	\includegraphics[width = 16cm]{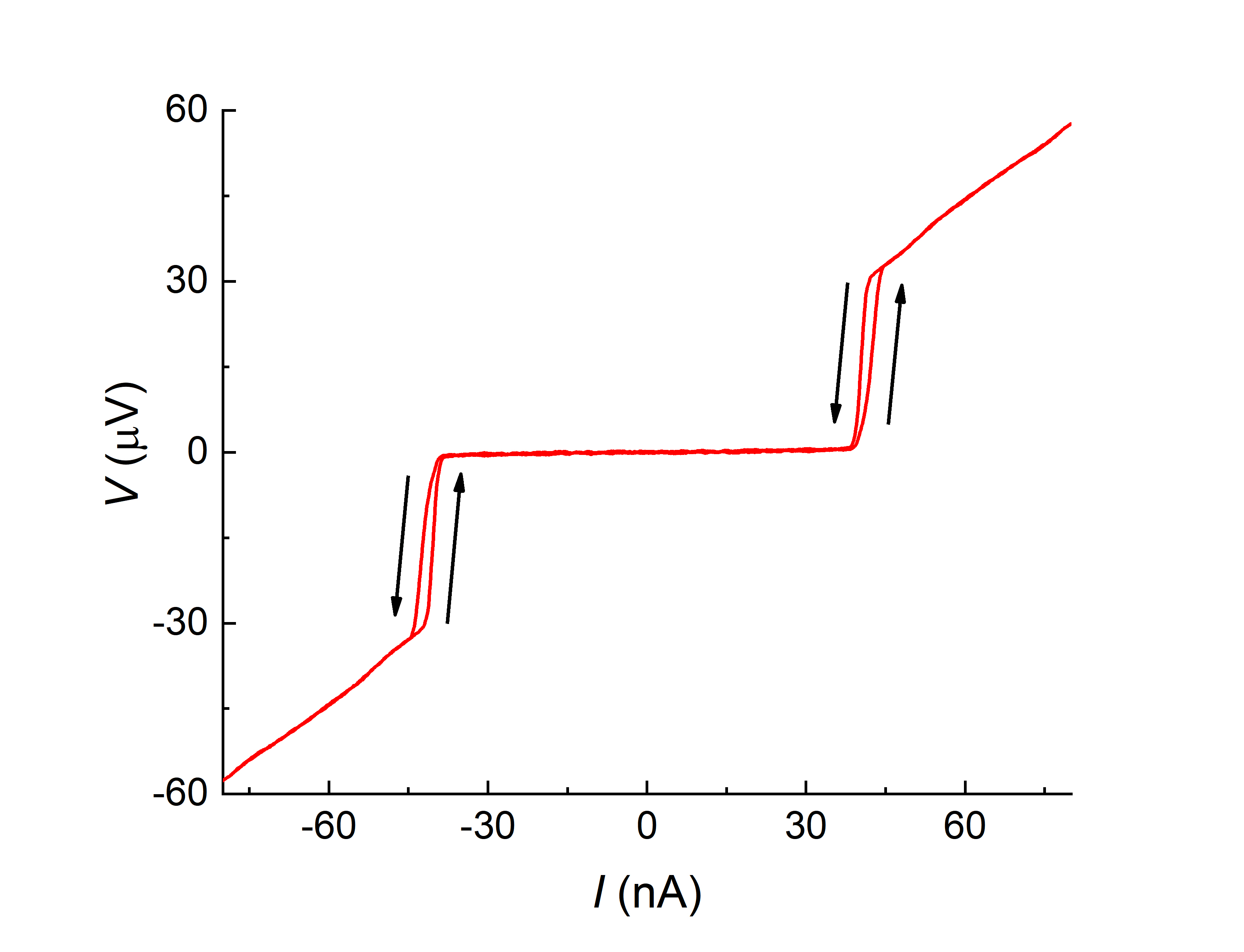}
	\centering
	\caption{$I(V)$ at $T=20$ mK. Arrows indicate the direction of current sweep.}
	\label{hyst}
\end{figure}

All the color maps R(I,H) presented in the paper was obtained in the following way: every V(I) curve of the raw data in V(I,H) was smoothed by 15 pts Adjacent-averaging, then dV/dI was numerically calculated and dV/dI(I) was smoothed again by the same procedure. 
The I-V curve, presented in the Fig.1(e) of the main text, is the average between 10 separately measured I-V curves. 
It was done to decrease the noise and to determine the resistance in the region between $I_{c1}$ and $I_{c2}$ more precisely. 
The critical current is determined as the maximum of $dV/dI(I)$, when the current is swept from zero to the maximum value.  
The retrapping current is determined as the maximum of $dV/dI(I)$, when the current is swept from the maximum value to zero. 
Errors are the width of the peak in $dV/dI(I)$. $I(V)$ curve with retrapping current is shown in the Fig.\ref{hyst}. 

Color map in the Fig.2(c) is not just zoom of the Fig.2(b), but a separate measurement.
Color map in the Fig.2(f) of the main text is slightly shifted (by 0.04 G) in order to get the maximum critical current at H=0. 
This shift is likely because of the trapped flux in the superconducting solenoid. 

We can estimate the number of channels from the geometry of our junction: 
\begin{equation}
    N=\frac{W}{\lambda_F/2}=\frac{500}{30/2}\approx30,
\end{equation}
where $W$ is the width of the Josephson junction, determined from SEM-Image, and $\lambda_F$ is the Fermi wavelength, determined from the ARPES spectra.
Also, we can estimate the number of conductive channels from the normal resistance, assuming that the interface between a superconductor and a topological insulator is transparent:
\begin{equation}
    N_{cond}=\frac{e^2/h}{R_N}=\frac{25812}{825}\approx31
\end{equation}
where $R_N$ is the normal resistance at higher currents. 
Those two values are consistent with each other. 
Nevertheless, as we will see next, not all of them are carrying supercurrent. Reasons for that are analyzed in details in [1].

From the critical temperature of SQUID we can estimate the proximity induced gap:
\begin{equation}
    \Delta=1.76k_BT_c=0.27\ meV,
\end{equation}
where $T_c$ is the critical temperature of the SQUID and $k_B$ is the Boltzmann constant. 
Of course, there is no evidence, that BCS theory works in such systems since it is predicted that p-wave superconductivity should exist in proximitized surface states of topological insulators. 
Nevertheless, measurements on similar systems show an induced gap of the same value.

Knowing the value of gap allows us to estimate the current, which is flowing through single conductive channel:
\begin{equation}
    I_Q=\frac{e\Delta}{2\hbar}=32\ nA.
\end{equation}

So, we can conclude that out of 30 channels, only a few of them are coherent. 
Nevertheless, this value does not coincide with the number of the channels from KO-2 fit (see the next paragraph).
Probably, this is because the BCS theory is not applicable here and the determination of the induced gap is not correct.

The period of the critical current oscillations is consistent with the SQUID loop area.
\begin{equation}
    \frac{\Phi_0}{\Delta H}=27.57 \ \mu m^2\approx5\times5\ \mu m^2,
\end{equation}
where $\Phi_0=\frac{h}{2e}$ is the superconducting flux quanta, $\Delta H$ is the period of the oscillations.

\begin{figure}[H]
	\includegraphics[width = 15cm]{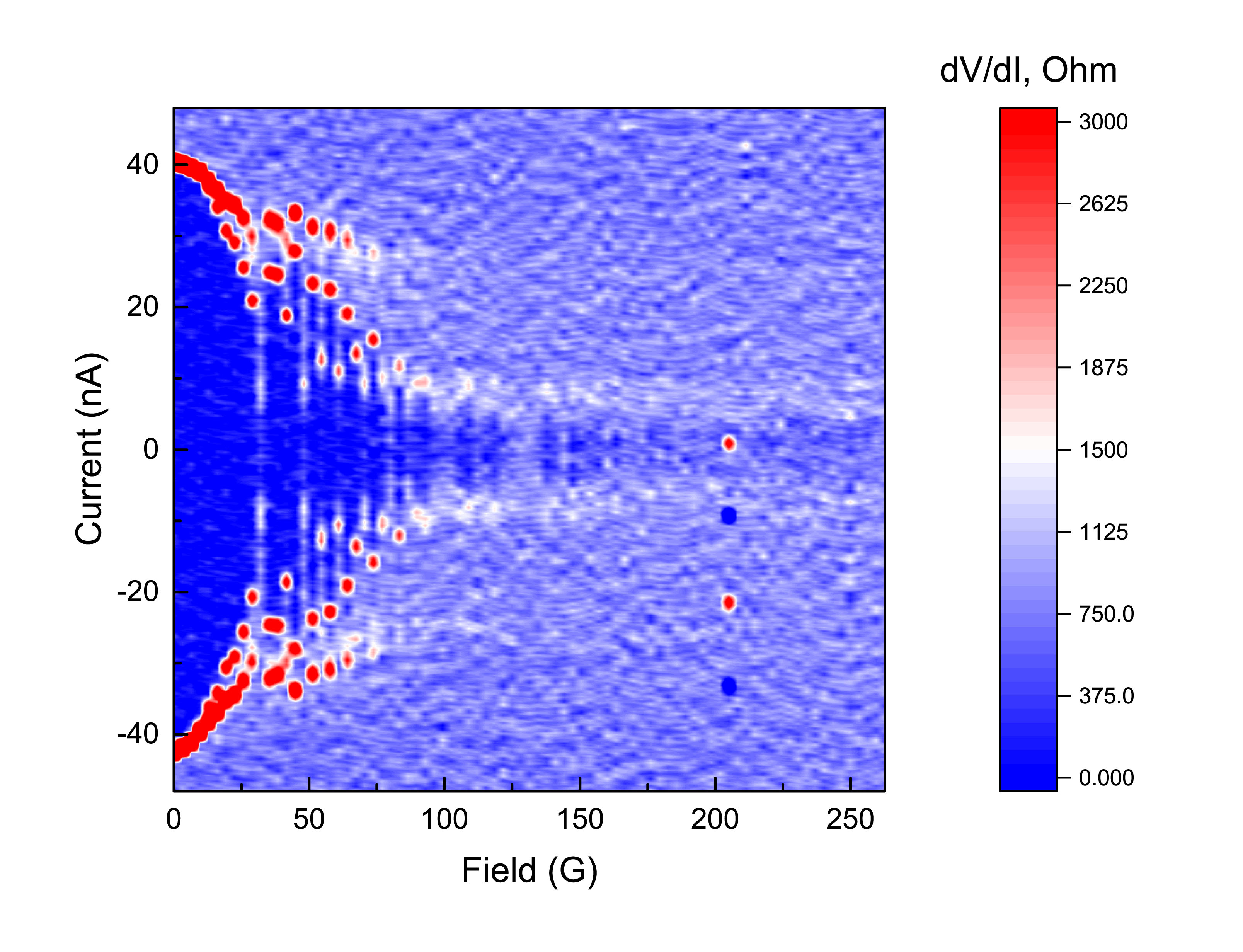}
	\centering
	\caption{Dependence of the critical current on external magnetic field  up to 250 G.}
	\label{250}
\end{figure}

The approximate area of the single Josephson junction is $0.1 um^2$, therefore the expected period of the envelope function of the $I_c(H)$ should be approximately 200 G. However, clearly, that Fig.\ref{250} does not show the Fraunhofer-like dependence. Moreover, from the Fig.2(b) one can conclude that the envelope function of the $I_{c1}(H)$ is concave, but the Fraunhofer-like function is convex.

\begin{figure}[H]
	\includegraphics[width = 15cm]{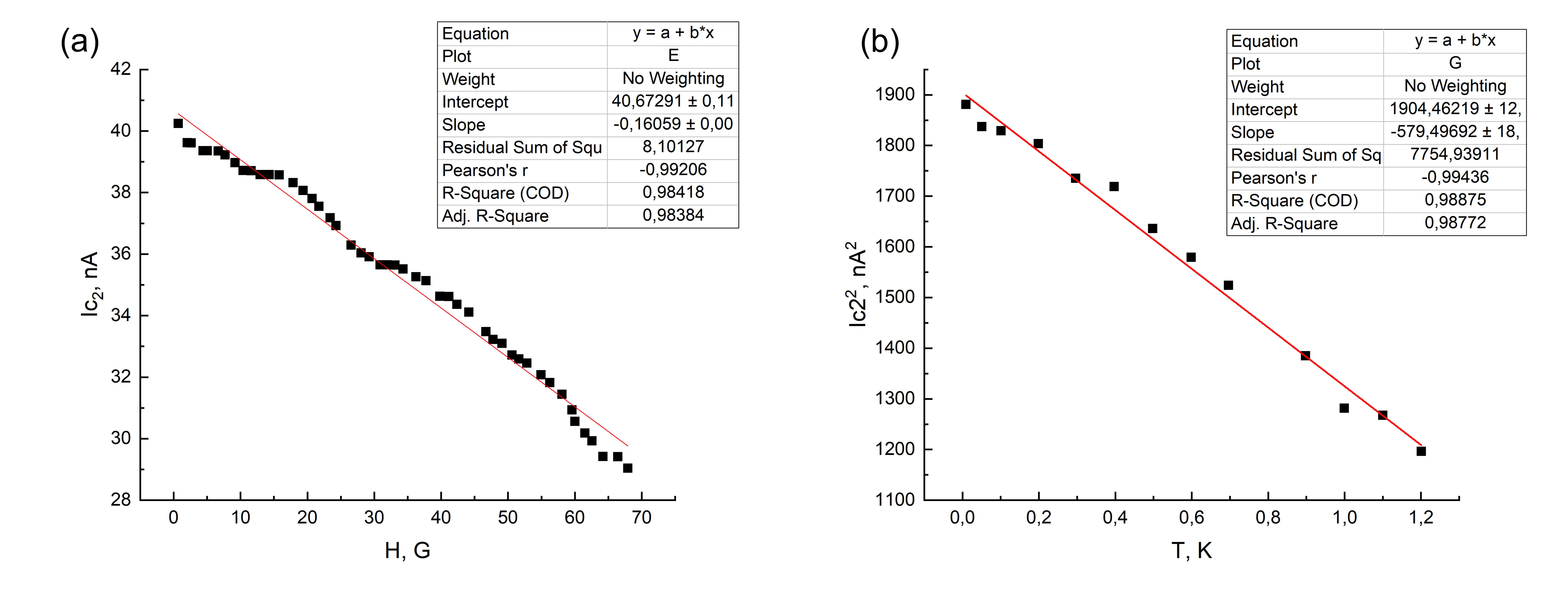}
	\centering
	\caption{(a) Linear fit of the $I_{c2}(H)$. (b) Linear fit of the $I^2_{c2}(T)$.}
	\label{Linear}
\end{figure}

Fig.\ref{Linear}(a) shows the linear dependence of the $I_{c2}$ on external magnetic field. 
To make squere-root dependence of the $I_{c2}$ more visible, we have plotted $I^2_{c2}(T)$ in Fig.\ref{Linear}(b).

\begin{figure}[H]
	\includegraphics[width = 16cm]{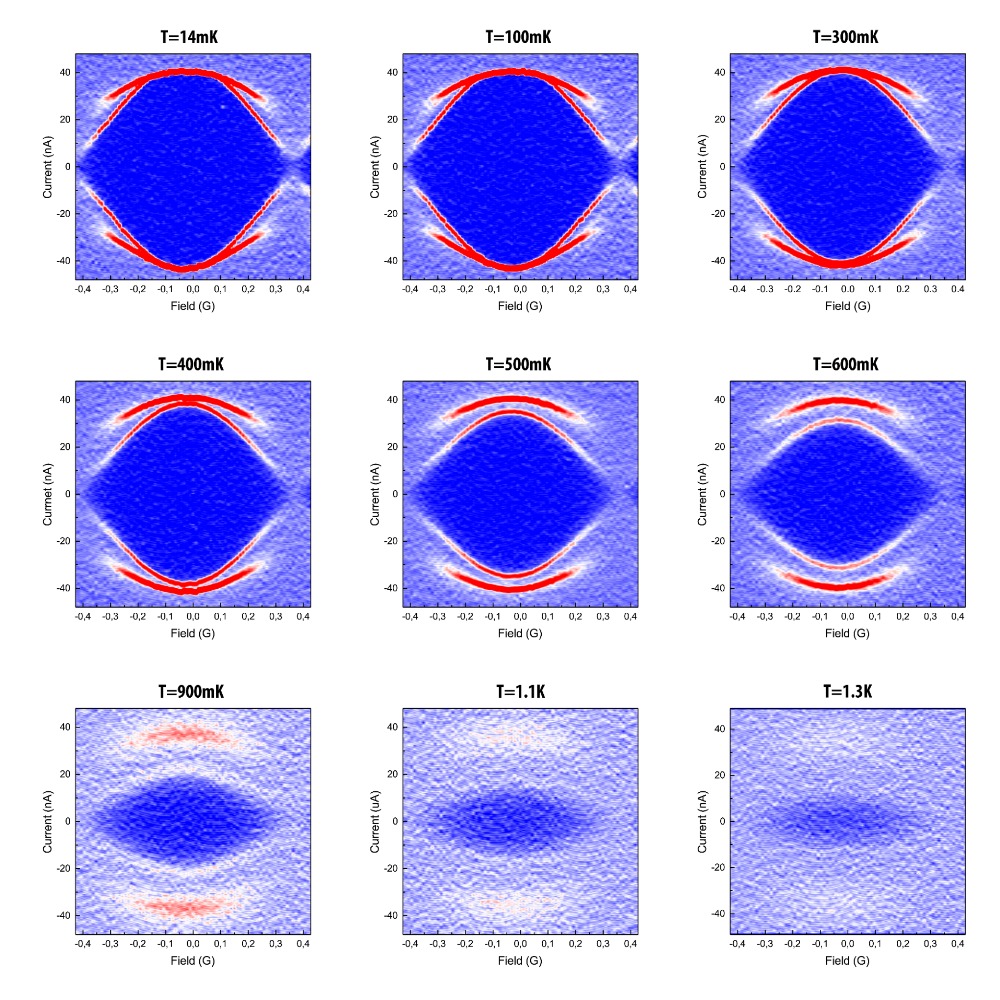}
	\centering
	\caption{Evolution of the R(I,H) on temperature. Red color correspond to the 2 kOhm resistance, blue - 0 Ohm. The current was swept from positive to negative values, therefore upper curve correspond to the retrapping current and lower curve - to the critical current.}
	\label{dataset}
\end{figure}

To make the situation with evolution of critical currents and retrapping current more transparent, we show here the set of R(I,H) color maps at different temperatures (see Fig.\ref{dataset}).

\section{Single Josephson junctions}
\begin{figure}[H]
	\includegraphics[width = 16cm]{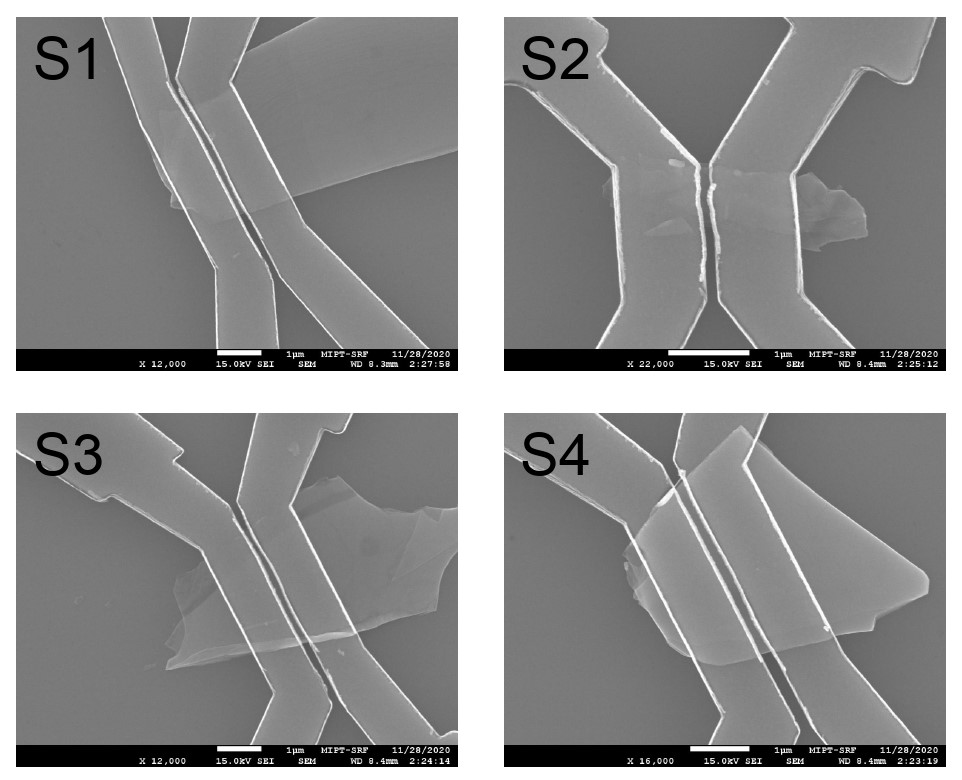}
	\centering
	\caption{SEM-Images of 4 single Josephson junctions, based on Fe-BTS flakes.}
	\label{BTS1}
\end{figure}

\begin{figure}[H]
	\includegraphics[width = 16cm]{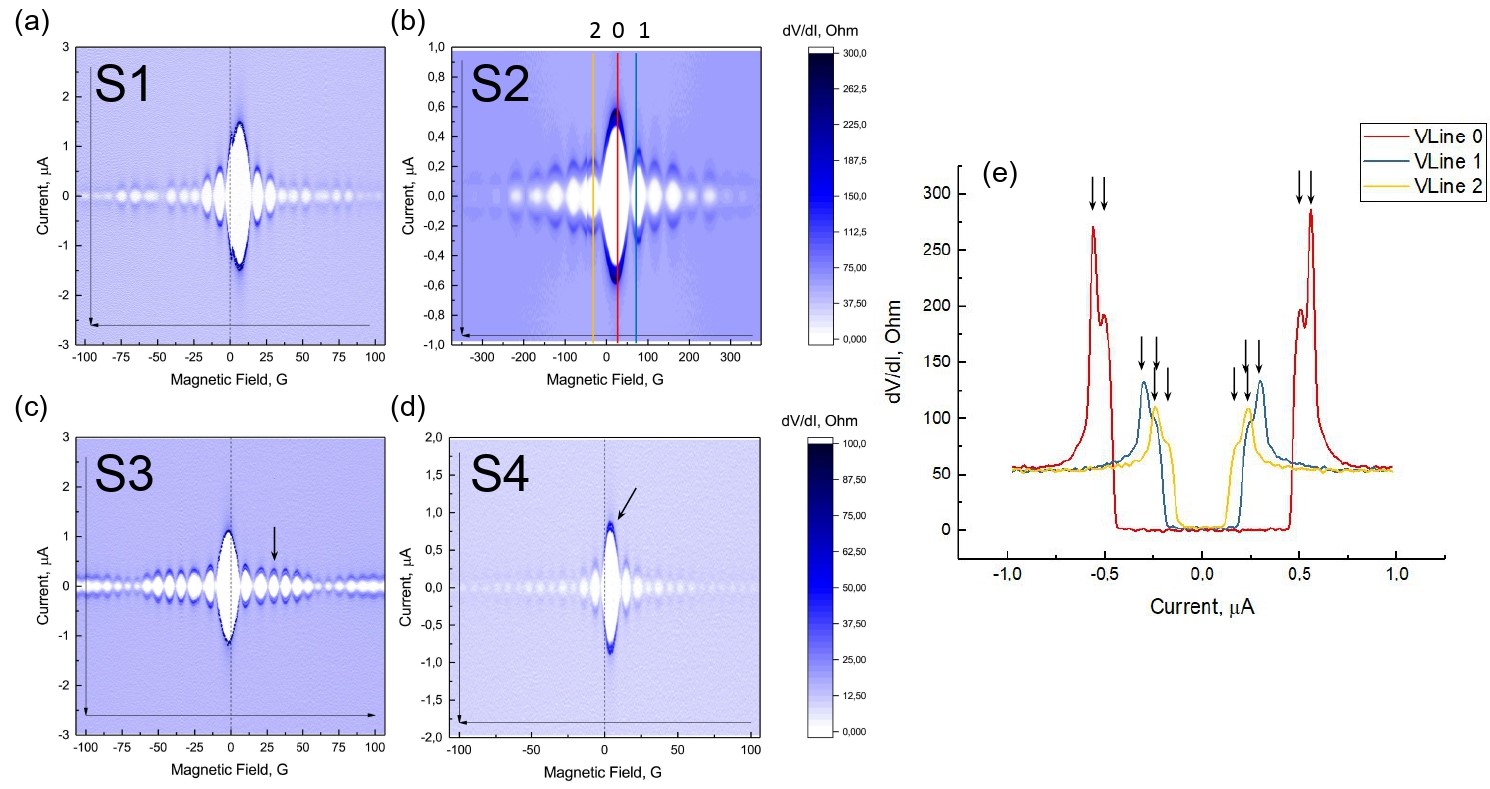}
	\centering
	\caption{(a)-(d) Resistance of 4 devices as a function of bias current and external magnetic field. Arrows point to the double critical current feature. (e) dV/dI(I) for S2 in magnetic fields, shown in (b) by color lines. Arrows indicate two critical currents.}
	\label{BTS2}
\end{figure}

Applying the same technological procedure, as described in the Section 1 of the Supplemental material, we have fabricated 4 single Josephson junctions based on Fe-BTS crystals [1]. SEM-Images of the final devices are shown in the Fig.\ref{BTS1}.
Three out of four devices (S2, S3, S4) show double critical current feature, as shown in the Fig.\ref{BTS2}. 
Moreover, the characteristic voltage, at with second step is observed is different in all three single JJ (S2, S3, S4) and the SQUID, which is shown in the main text. 

\section{Estimation of transparency of the interface}
The detailed information about the KO-2 model can be found in the  Methods section of our previous work [2]. 
In order to fit the $I_{c1}(T)$ we used the least square method with 3 variables: critical temperature of the Josephson junction $T_c$, the number of channels N and transparency of the NS interface D. 
Width W and length L of the junction was determined from the SEM and TEM, respectively. 
Fermi velocity $v_F$ and the Fermi wavelength $\lambda_F$ was determined from the ARPES. It should be noticed, that the critical current of single Josephson junction is  2 times smaller, then critical current of SQUID (assuming the high symmetry of the SQUID). 
All the values for single Josephson junction are presented in the Table \ref{table:1}. As one can see, the determined transparency is 98\%.

\begin{table}[h!]
\centering
\begin{tabular}{||c c c c||} 
 \hline
 Parameter & Value & Standard Error & Vary? \\ [0.5ex] 
 \hline\hline
 $T_c$  & 1.81 K & 2.41\% & Yes \\ 
 $v_F$  & $4.6\times10^5$ m/s & - & No \\
 W & $5\times10^{-7}$ m & - & No \\
 L & $1.4\times10^{-7}$ m & - & No \\
$\lambda_F$  & $3\times10^{-8}$ m & - & No \\
 N & 7.1 & 5.33\% & Yes \\
 D & 0.981 & 0.53\% & Yes \\ [1ex] 
 \hline
\end{tabular}
\caption{Results of fitting the $I_{c1}(T)$ by the KO-2 model.}
\label{table:1}
\end{table}

\begin{figure}[H]
	\includegraphics[width = 16cm]{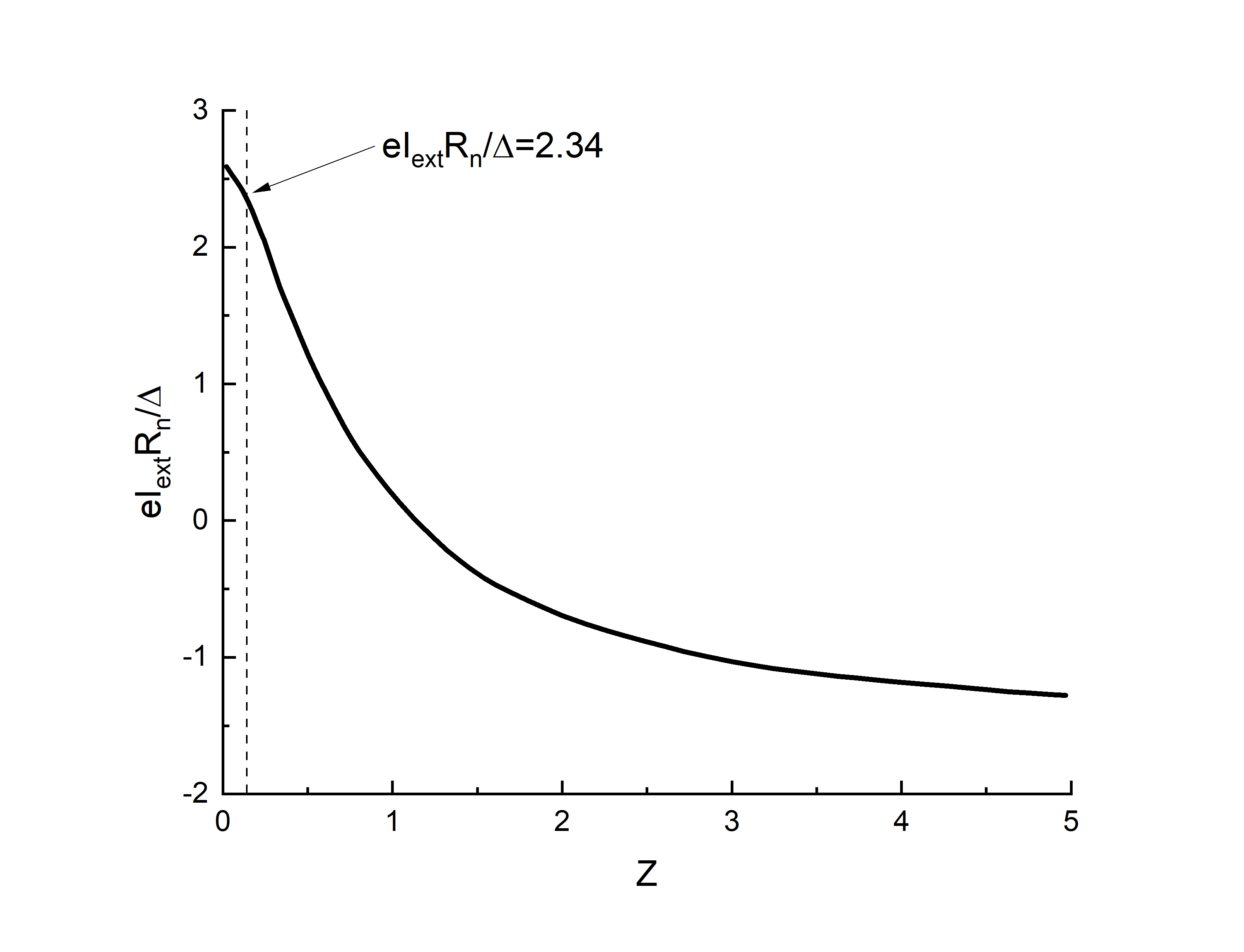}
	\centering
	\caption{Normalized excess current as a function of
the scattering parameter Z, according to [4]. Dashed line shows the value of Z, which corresponds to transparency of 98\%.}
	\label{transp}
\end{figure}

Other way to determine transparency is using Octavio-Tinkham-Blonder-Klapwijk theory[4]. 
In this approach, one have to use the value of the excess current, the normal state resistance and the value of the induced superconducting gap to calculate the scattering parameter Z, which is related to the transparency. 
The problem in our case is that we do not know the exact value of the gap, which is typically determined via multiple Andreev reflections. 
But we can estimate the value of the gap, using transparency, obtained via KO-2 fit of the first critical current. 
In this case, the normalized excess current should be 2.34 (see Fig.\ref{transp}), which correspond to $\Delta=0.047$mV. 
This value does not match the value obtained via $T_c$ and typical values, observed in similar system.
If we take the value of the gap $\Delta=0.27$mV, the transparency is 57\%. 
It may be explained, taking into account the complex structure of the interface. Probably, KO-2 model shows the transparency of the $S^{\prime}N$ interface because it takes into account only the $I_{c1}(T)$ dependence, and Octavio-Tinkham-Blonder-Klapwijk theory shows the average transparency between two superconducting electrodes. Nevertheless, the applicability of both theories to such complex interfaces is questionable. 

\section{RSJ-Model}
\begin{figure}[H]
	\includegraphics[width = 16cm]{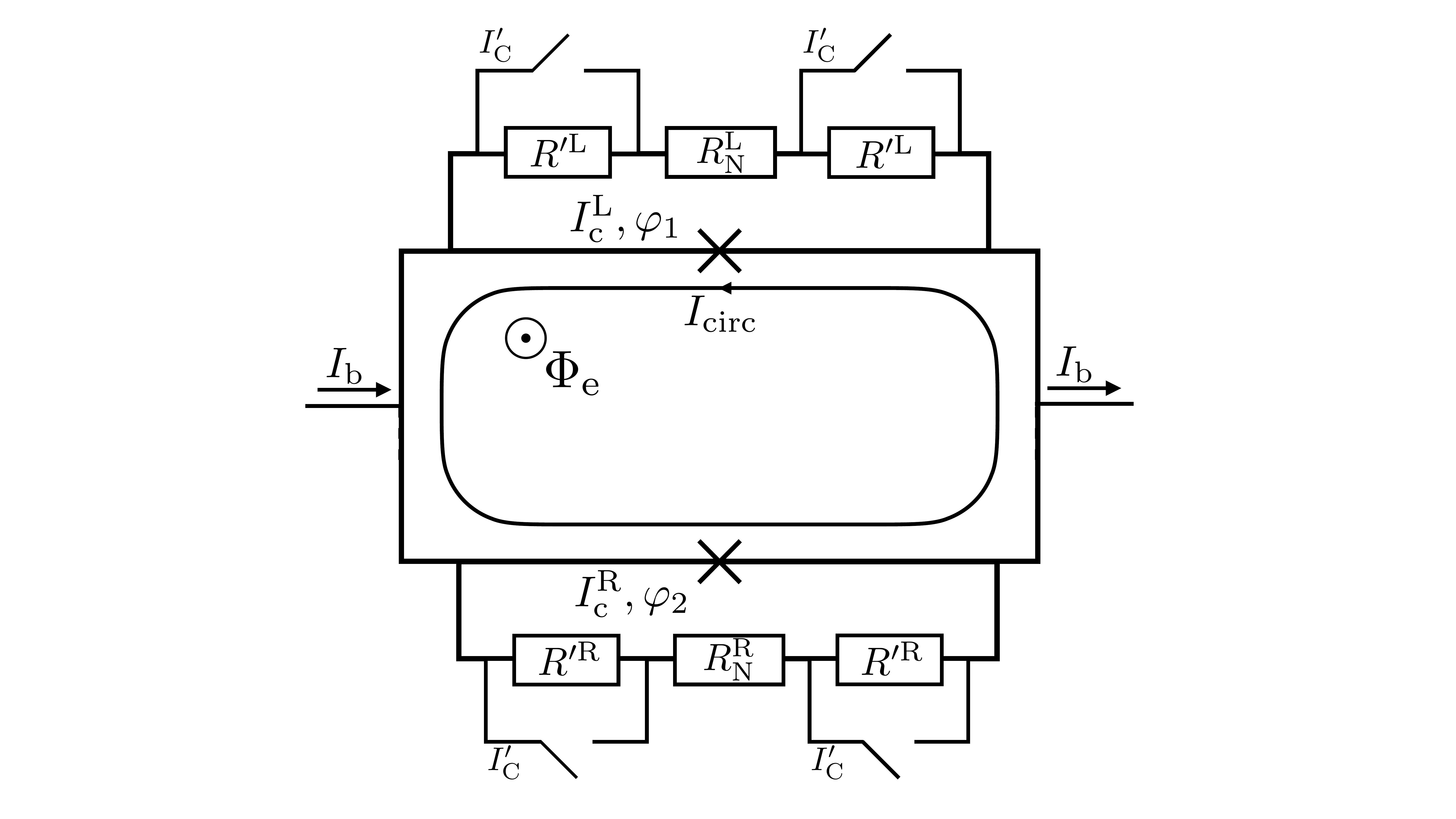}
	\centering
	\caption{Equivalent scheme for the SQUID.}
	\label{schm}
\end{figure}

We introduce the following modification of the RSJ model (equivalent scheme is presented in Fig. \ref{schm}).
We model two interfaces in each junction as  thin films that have  critical currents $I_{c}$ and normal resistances  $R'^{R,L}$. 
Resistances of the middle part of the junctions are $R_N^{R,L}$.

There are at least two ways to introduce the switching condition of the superconducting interface to normal state.
The first is just $|I^L|>I_c \; \text{and}\; |I^R|>I_c$ and the second one is $|I_b/2|+|I_{circ}|> I_{c}$. 
The second condition may be interpreted as the Cooper pair density saturation condition. 
This is valid only for junctions in symmetrical SQUID, so we assume that critical currents of all interfaces are equal when the second condition is applied. 

\begin{figure}[H]
	\includegraphics[width = 0.49\textwidth]{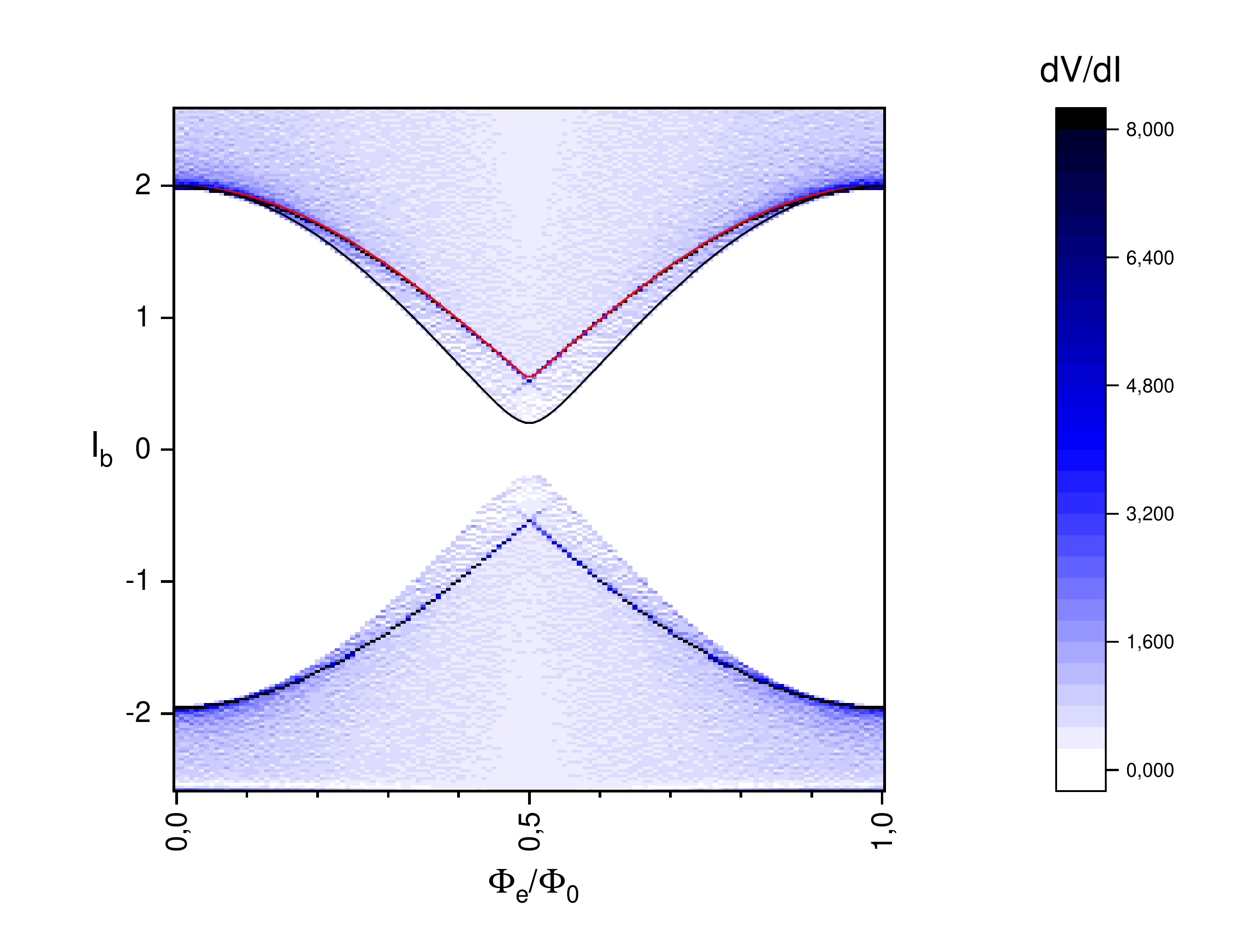}
		\includegraphics[width = 0.49\textwidth]{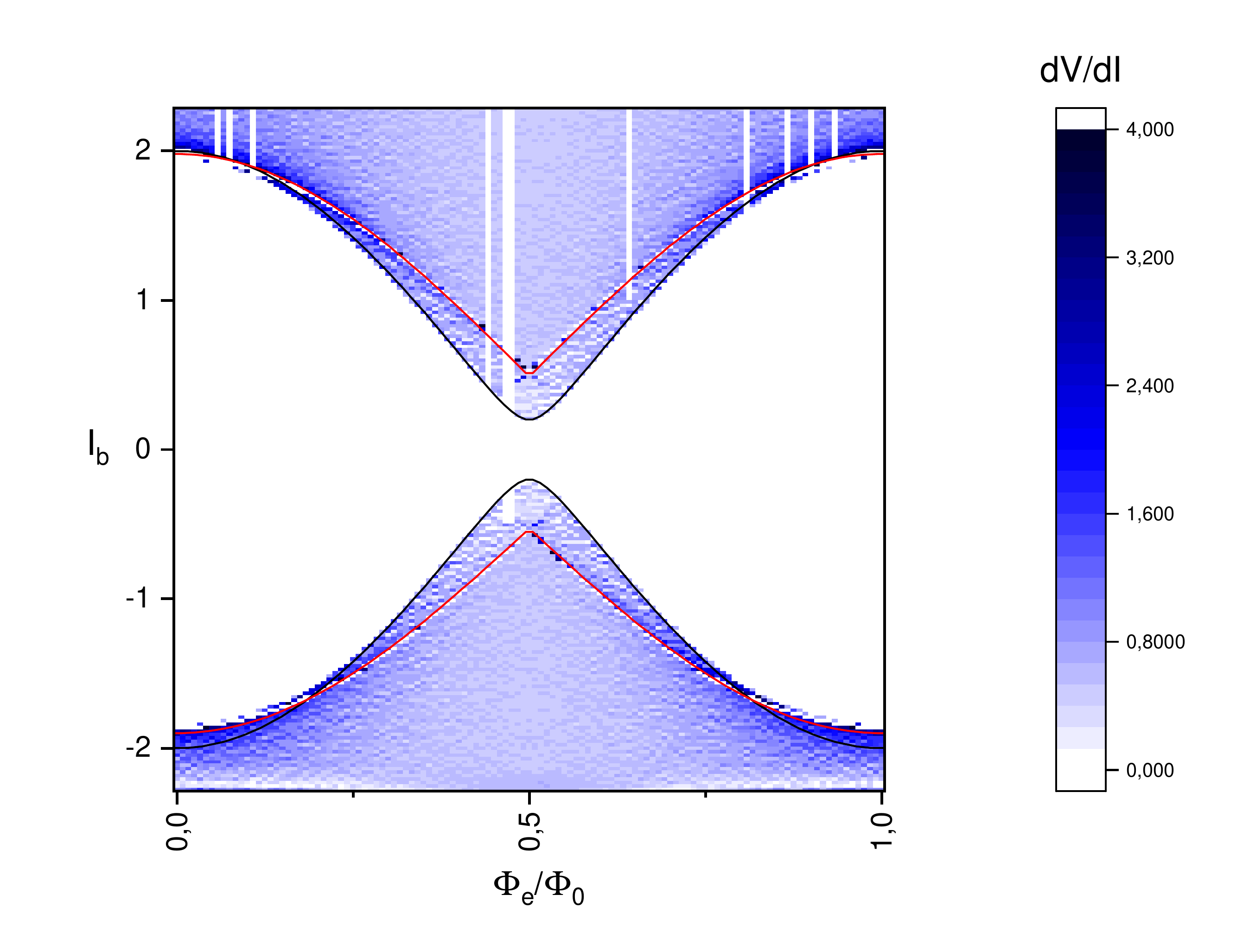}
	\centering
	\caption{Results of numerical simulation according to simplified model with the second switching condition.  Joule heat added to the system on the right figure. }
	\label{map}
\end{figure}

The introduced model suggests the following equations:
\begin{equation}
\begin{gathered}
\begin{cases}
		I_b = i^R+i^L;\\
		I_{circ} = \frac{(i^R-i^L)}{2};\\
		i^{L,R} =I_{c}^{L,R}\sin{\varphi_{1,2}}+\frac{\dot{\varphi_{1,2}}}{R_{N}^{L,R}+2R'^{L,R}(i^L,i^R)};\\
		R'^{L,R}(i^L,i^R) =R'^{L,R}\cdot\operatorname{\theta}\left( \bigg|\frac{I_b}{2}\bigg|+|I_{circ}|-I_{c}\right);  \\
		\varphi_1-\varphi_2 = 2\pi\Phi_e/\Phi_0.
		\end{cases}
\end{gathered}
\end{equation}
Here we have used dimensionless variables according to [5] and the second switching condition as an example.
For numerical simulation we used slightly simplified equations:
\begin{equation}
	\begin{gathered}
	\begin{cases}
		I_b = i^R+i^L;\\
			i^{L,R} =I_{c}^{L,R}\sin{\varphi_{1,2}}+\frac{\dot{\varphi_{1,2}}}{R_{N}^{L,R}+2R'^{L,R}(i^L,i^R)};\\
				i^{*L,R} =I_{c}^{L,R}\sin{\varphi_{1,2}}+\frac{\dot{\varphi_{1,2}}}{R_{N}^{L,R}};\\
		R'^{L,R}(i^L,i^R) =R'^{L,R}\cdot\operatorname{\theta}\left( \bigg|\frac{i^{*L}+i^{*R}}{2}\bigg|+\bigg|\frac{i^{*L}-i^{*R}}{2}\bigg|-I_{c}\right);  \\
		\varphi_1-\varphi_2 = 2\pi\Phi_e/\Phi_0.
		\end{cases}
	\end{gathered}
\end{equation}
The solutions of the simplified system of equations (Fig. \ref{map}) are in good agreement with the experiment, that is why we consider the simplified model to be applicable.
The second "critical" current coincides with $A+B|\cos(\Phi/\Phi_0)|$ in both numerical (Fig. \ref{map}) and experimental results, where $A$ and $B$ are fitting parameters. 
Additionally, we observed that the amplitude of the second voltage step correlates with N' resistance value, as expected.
Matching numerical results and the experiment we found that $R'\sim R_N$.

The experimental results show hysteresis and blurriness of the second voltage step in half flux quantum region. 
To explain the observed properties of the system we successively add new mechanisms to the model. 

To observe hysteresis we added Joule heating by introducing equations of heat balance and allowing components of the JJ to transfer heat across the boundaries according to the Newton's law. 
If we say that the whole JJ is at the same temperature then we will not get any hysteresis. 
The desired effect is observed only when we introduce the temperatures of the interfaces. 
In both the model and the experiment the retrapping current limits $I_{c1}$. 
In other words, retrapping current appears heated whereas the first critical current does not, which can be seen from the corresponding fits  (Fig. \ref{map}). 

\begin{figure}[H]
	\includegraphics[width = 0.6\textwidth]{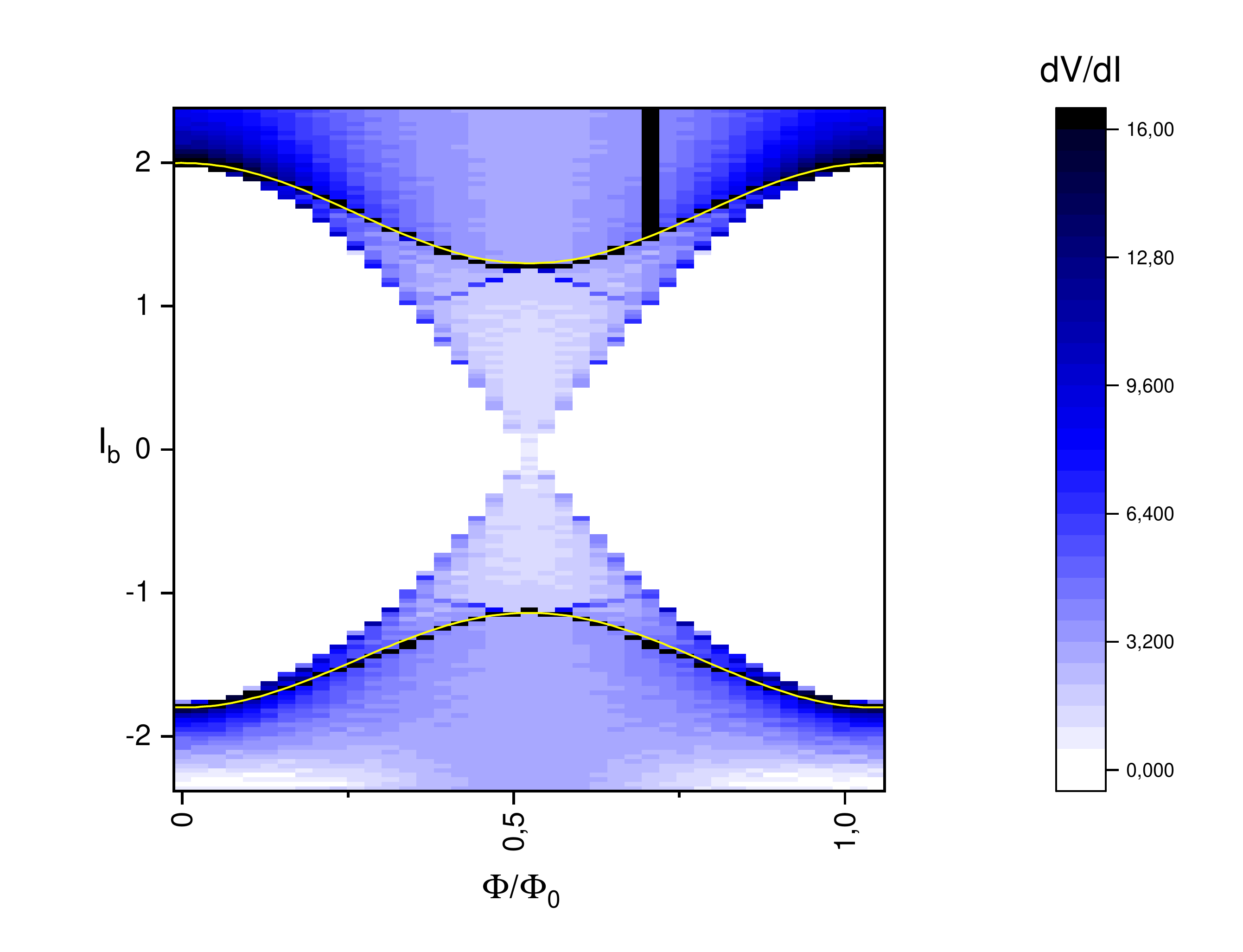}
	\centering
	\caption{Results of numerical simulation with the first condition applied.}
	\label{normmap}
\end{figure}

We introduced multilayered interface to observe blurriness.
Each layer should produce a small voltage step, near integer flux quantum regions steps should lay closer and form a bigger step. 
Approaching the half integer flux quantum regions, small steps should slide apart and create the blurriness effect. 
However, the numerical model exhibits complex behaviour, e.g.
the desired effect is observed only in the case of $R'$ resistance asymmetry relatively to $R_N$ resistance value.
Additionally, we observed that $R'\sim R_{N}$ should hold for visible blurriness. 

If the first condition is applied, the second critical current is $A+B\cos(2\pi\Phi/\Phi_0)$, according to the fit (Fig. \ref{normmap}). 
Note that the shape of the $|\cos(x)|$ and the $\cos(x)$ coincide in the origin of the maximum with proper amplitude tuning, therefore both conditions provide great fits of the experimental data, because for both switching conditions obtained fits are similar in the origin of integer flux quantum and $I_{c2}$ does not extend enough in between to observe the deviation. 
However, the hysteretic retrapping current in the experiment can not be fitted with cosine: it extends enough to see the mismatch of the fit and experiment. 
This suggests that the second switching condition is more appropriate.